\documentclass[12pt]{article}
\usepackage{graphicx}
\def \bea{\begin{eqnarray}}
\def \beq{\begin{equation}}

\def \eea{\end{eqnarray}}
\def \eeq{\end{equation}}

\def \ket#1{| #1 \rangle}

\def \od{\overline{D}^0}

\def \pp{\psi'}
\def \ppp{\psi''}

\begin{document}
\rightline{EFI 04-38}
\rightline{hep-ph/0411003}
\bigskip
\centerline {\bf $\ppp$ DECAYS TO CHARMLESS FINAL STATES
\footnote{To be published in Ann.\ Phys.}}
\bigskip
 
\centerline{Jonathan L. Rosner~\footnote{rosner@hep.uchicago.edu.}}
\centerline{Enrico Fermi Institute and Department of Physics}
\centerline{University of Chicago, 5640 S. Ellis Avenue, Chicago, IL 60637}
\medskip
 
\begin{quote}

The importance of measuring the non-$D \bar D$ decays of the $\ppp =
\psi(3770)$ resonance is discussed.  These decays can shed light on a possible
discrepancy between the total and $D \bar D$ cross sections at the $\ppp$, and
on a proposed mechanism for enhancement of penguin amplitudes in $B$ meson
decays through charm--anticharm annihilation.
Measurements (including the $\psi''$ line shape) in states of definite G-parity
and in inclusive charmless final states such as $\eta' + X$ are found to be
particularly important.

\end{quote}

\bigskip

The $\ppp \equiv \psi(3770)$ particle,\footnote{Numbers in parentheses denote
masses in MeV$/c^2$} lying just above $D \bar D$ threshold, is a well-defined
source of charmed particle pairs in $e^+ e^-$ collisions.  It is now
undergoing high-intensity studies at the CLEO Detector at Cornell \cite{CLEOc}
and the BES Detector in China \cite{Ablikim:2004ck}.  Its
couplings to charmless states are of interest for several reasons.

(1) The production and decays of $\ppp$ depend on its composition of $^3D_1$,
$^3S_1$, and $D \bar D$ continuum states \cite{Eichten:2002qv,Eichten:2004uh}.
Mixing among these states also can affect $\psi'$ modes, suppressing some while
leading to contributions in $\ppp$ decays \cite{Rosner:2001nm}.  These effects
can be subtle as a result of interference \cite{Wang:2003hy,Wang:2004kf}.

(2) New measurements of $\sigma(e^+ e^- \to \ppp \to D \bar D) \equiv \sigma(D
\bar D)$ by BES \cite{Ablikim:2004ck} and CLEO \cite{He:2005bs} confirm
an earlier
result \cite{Adler:1987as} that $\sigma(D \bar D)$ is less than the total cross
section $\sigma(e^+ e^- \to \ppp \to \ldots) \equiv \sigma(\ppp)$ measured by
several groups \cite{Partridge:1984kb,Peruzzi:1977ms,Schindler:1980ws,%
Bai:2001ct}.  The ratio $\sigma(D \bar D)/\sigma(\ppp)$ is of intrinsic
interest and provides an estimate for rates for channels other than $D \bar D$
during forthcoming extensive accumulations of data at the $\ppp$ energy.

(3) The non-charm decays of $\ppp$, if appreciable, provide a possible
laboratory for the study of rescattering effects relevant to $B$ meson decays.
If the $\ppp$ decays to $D \bar D$ pairs which subsequently re-annihilate into
non-charmed final states, similar effects can generate enhanced penguin
amplitudes (particularly in $b \to s$ transitions) in $B$ decays.
Re-annihilation mechanisms are relevant not only for heavy quarkonium decays
into non-flavored final states \cite{Lipkin:1986av,Achasov:vh} but also for
non-$K \bar K$ decays of the $\phi$ meson \cite{Achasov:1999tp}.

Non-charmed final states of the $\ppp$ were discussed in doctoral theses
\cite{Zhu:1988,Majid:1993} based on Mark III data.  No significant signals were
obtained.  While the total width of $\ppp$ is $\Gamma(\ppp) = 23.6 \pm 2.7$
MeV \cite{PDG}, partial widths to $\gamma \chi_{cJ}~ (J=1,2)$ are expected not
to exceed a few tens of keV, with a few hundred keV expected for $\Gamma(\ppp
\to \gamma \chi_{c0})$.  The partial width $\ppp \to \pi \pi J/\psi$ is not
expected to exceed about 100 keV.  Thus any non-$D \bar D$ branching ratio in
excess of $\sim 2\%$ must come from as-yet-unseen channels.

In this article I discuss the known $\ppp$ decays, including $D \bar D$, lepton
pairs, $\gamma \chi_{cJ}$, and $J/\psi \pi^+ \pi^-$, noting the likelihood of
an appreciable non-$D \bar D$ cross section.  A model for this contribution due
to re-annihilation of $D \bar D$ pairs into light quarks is presented.  It
implies signatures from interference with the continuum process $e^+ e^- \to
\gamma^* \to {\rm~light}~q \bar q {\rm~pairs}$.  {\it Inclusive} measurements
to states with definite G-parity then become useful, and charmless $\ppp$
decays can illuminate some classes of $B$ decays including those with $\eta'$.
Rates for $\ppp$ decays to specific charmless final states have recently been
estimated in Ref.\ \cite{Wang:2004kf}. 

One measures $D \bar D$ production at the $\ppp$ by comparing the rates for
$e^+ e^- \to \ppp \to f_i + \ldots$ and $e^+ e^- \to \ppp \to f_i \bar f_j$,
where $f_i$ and $f_j$ are final states in $D$ decay.  Unknown branching ratios
can be determined, but one must know detector efficiency well.  This method was
used by the Mark III Collaboration \cite{Adler:1987as} with an integrated
luminosity $\int {\cal L} dt = (9.56 \pm 0.48)$ pb$^{-1}$.  The CLEO
Collaboration measured $\sigma(D \bar D)$ using this method with $\int {\cal L}
dt \simeq 56$ pb$^{-1}$ \cite{He:2005bs}.  The values are compared with those
from Mark III and the BES Collaboration \cite{Ablikim:2004ck} (with $\int {\cal
L} dt = 17.7$ pb$^{-1}$) in Table \ref{tab:sigDD}.

\begin{table}
\caption{Comparison of cross sections $\sigma(D \bar D) \equiv \sigma(e^+ e^-
\to \ppp \to D \bar D)$, in nb.
\label{tab:sigDD}}
\begin{center}
\begin{tabular}{l c c c} \hline \hline
Collaboration & $\sigma(D^+ D^-)$ & $\sigma(D^0 \bar D^0)$ & $\sigma(D \bar D)$
 \\ \hline
BES-II \cite{Ablikim:2004ck} & $2.56 \pm 0.08 \pm 0.26$ & $3.58 \pm 0.09
 \pm 0.31$ & $6.14 \pm 0.12 \pm 0.50$ \\
CLEO \cite{He:2005bs} &$2.79\pm0.07^{+0.10}_{-0.04}$
 & $3.60\pm0.07^{+0.07}_{-0.05}$ & $6.39 \pm 0.10^{+0.17}_{-0.08}$ \\
Mark III \cite{Adler:1987as} & $2.1 \pm 0.3$ & $2.9 \pm 0.4$ & $5.0 \pm 0.5$ \\
\hline \hline
\end{tabular}
\end{center}
\end{table}

The ratios $\sigma(D^+ D^-)/\sigma(D^0 \bar D^0)$ are consistent with the ratio
$(p^*_{+-}/p^*_{00})^3 = 0.685$ appropriate for the P-wave decay $\ppp \to D
\bar D$ (where $p^*$ denotes the magnitude of the center-of-mass [c.m.]
3-momentum).  Coulomb and other final-state-interaction effects can alter this
ratio and lead to its dependence on energy \cite{Voloshin:2004nu}.

\begin{table}
\caption{Comparison of total cross sections $\sigma(\ppp) \equiv \sigma(e^+ e^-
\to \ppp \to \ldots)$, in nb.
\label{tab:sig}}
\begin{center}
\begin{tabular}{l c} \hline \hline
Collaboration & $\sigma(\ppp)$ \\ \hline
Crystal Ball \cite{Partridge:1984kb}  & $6.7 \pm 0.9$  \\
Lead-Glass Wall \cite{Peruzzi:1977ms} & $10.3 \pm 1.6$ \\
Mark II \cite{Schindler:1980ws}       & $9.3 \pm 1.4$  \\
BES$^a$ \cite{Bai:2001ct}            & $7.7 \pm 1.1$  \\ \hline
Average                               & $7.9 \pm 0.6$  \\ \hline \hline
$^a$ Estimate based on fit (see text). & \\
\end{tabular}
\end{center}
\end{table}

\begin{figure}
\includegraphics[width=\textwidth]{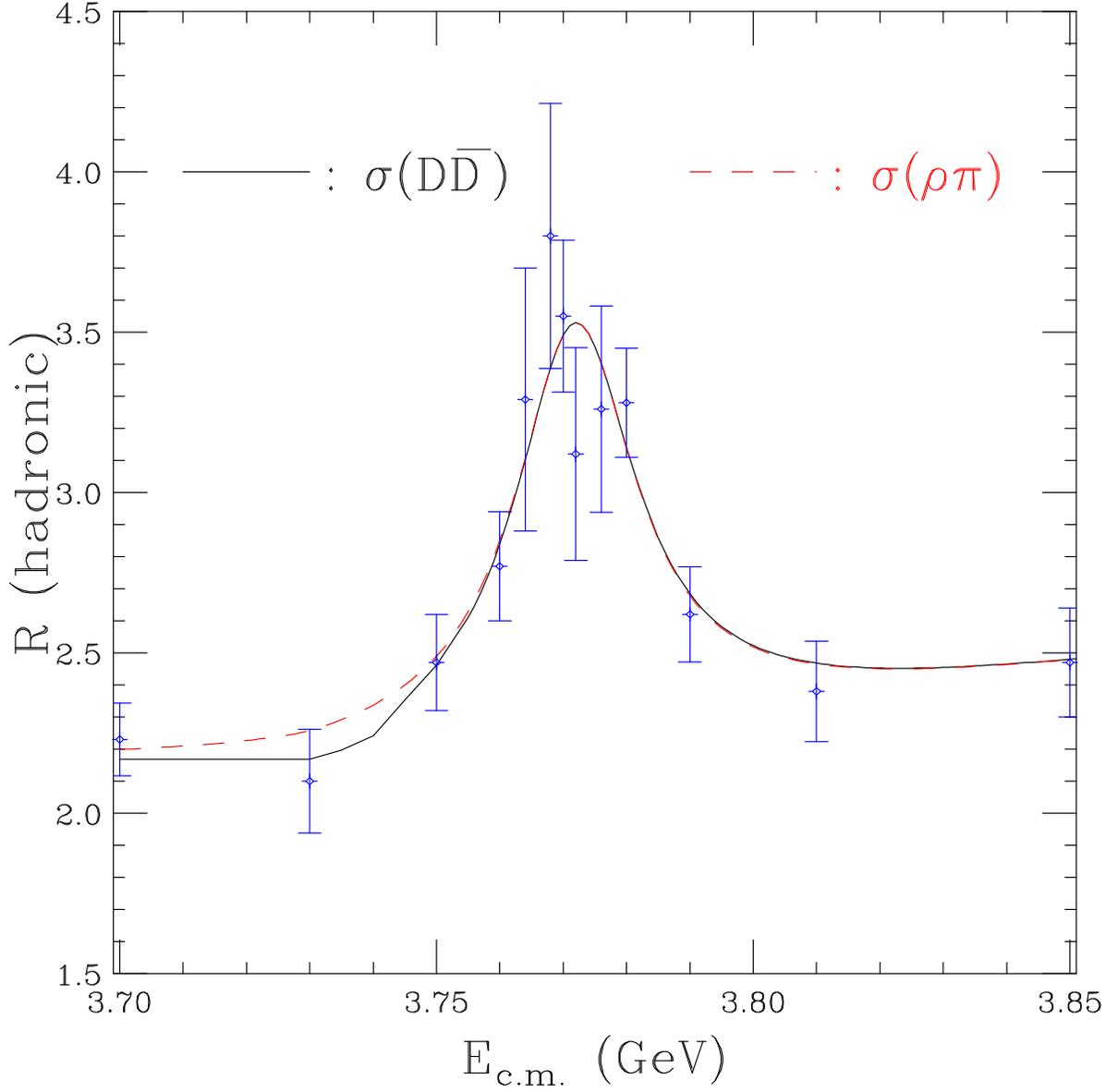}
\caption{Fit to the $\ppp$ peak in BES data \cite{Bai:2001ct}.  Solid line
denotes expected line shape for a $D \bar D$ final state, incorporating
appropriate centrifugal barrier terms, while dashed line denotes expected
line shape for $\rho \pi$ final state.
\label{fig:sigppp}}
\end{figure}

The values in Table \ref{tab:sigDD} are to be compared with those for the
total cross section $\sigma(\ppp)$ in Table \ref{tab:sig}.
In Fig.\ \ref{fig:sigppp} the BES data \cite{Bai:2001ct} on $R = \sigma(e^+ e^-
\to {\rm hadrons})/\sigma(e^+ e^- \to \mu^+ \mu^-)$ are displayed, along with
the results of a fit to the resonance shape using conventional Blatt-Weisskopf
angular momentum barrier factors \cite{BW}.  The fit obtains
$\sigma_{\rm pk} = 7.7 \pm 1.1$ nb, with other central values $M = 3772$ MeV,
$\Gamma = 23.2$ MeV, and $R_{\rm bg} = 2.17 + 2.36(E_{\rm c.m.} - 3.73~{\rm
GeV})\theta(E_{\rm c.m.} - 3.73~{\rm GeV})$.  The threshold energy 
is held fixed \cite{TSPC}.

It appears that $\sigma(D \bar D)$ falls short by one or more nb from the total
cross section $\sigma(\ppp)$.  Improved measurements of both quantities by the
{\it same} experiment will be needed to resolve the question.  I will show the
effect of ascribing this difference to $D \bar D$ re-annihilation into
light-quark states.  First I discuss other non-$D \bar D$ final states of
$\ppp$.

\begin{figure}
\begin{center}
\includegraphics[height=6.5in]{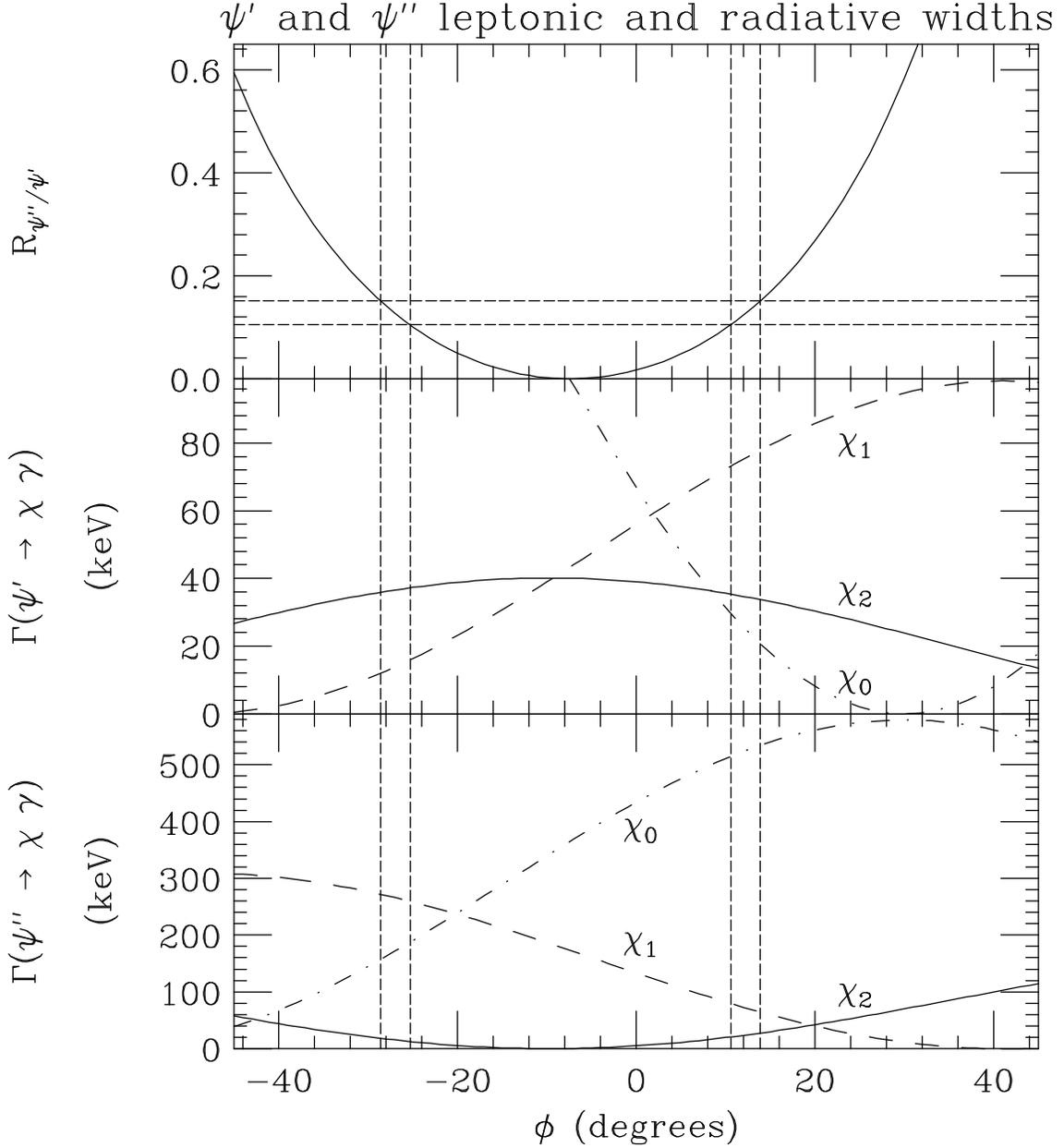}
\end{center}
\caption{Sensitivity of scaled leptonic width ratio $R_{\ppp/\psi'}$ and
partial widths $\Gamma(\psi',\ppp \to \chi \gamma)$ to mixing angle $\phi$.
Horizontal lines in top panel denote $\pm 1 \sigma$ limits on $R_{\ppp/\psi'}$,
and are projected onto the $\phi$ axis with vertical bands.  In middle and
bottom panels solid, dashed, and dash-dotted curves denote partial widths to
$\gamma \chi_{c2}$, $\gamma \chi_{c1}$, and $\gamma \chi_{c0}$, respectively.
\label{fig:psimix}}
\end{figure}

The leptonic width $\Gamma(\ppp \to e^+ e^-)$ is $0.26 \pm 0.04$ keV
\cite{PDG}, about 1/8 that of $\psi'$.
A simple model of S--D wave mixing for the $\psi'$ and $\ppp$ is to write
\begin{equation}\label{eqn:mix}
\ppp = \cos \phi \ket{1 ^3D_1} + \sin \phi \ket{2 ^3S_1}~~,~~~
\psi' = -\sin \phi \ket{1 ^3D_1} + \cos \phi \ket{2 ^3S_1}~~.
\end{equation}
The ratio $R_{\ppp/\psi'}$ of leptonic widths (scaled by factors of $M^2$) and
the partial widths $\Gamma(\psi' \to \chi \gamma)$ and $\Gamma(\ppp \to \chi
\gamma)$ may then be calculated as functions of $\phi$ \cite{Rosner:2001nm,%
Kuang:2002hz}.  Specifically, it was found in Ref.\ \cite{Rosner:2001nm} that

\begin{equation}
R_{\ppp/\psi'} \equiv \frac{M_{\ppp}^2 \Gamma(\ppp \to e^+ e^-)}
{M_{\pp}^2 \Gamma(\pp \to e^+ e^-)}
= \left| \frac{0.734 \sin \phi + 0.095 \cos \phi}
              {0.734 \cos \phi - 0.095 \sin \phi} \right|^2
= 0.128 \pm 0.023~~,
\end{equation}
while
\begin{equation}
\Gamma(\ppp \to \gamma \chi_{c0}) = 145~{\rm keV} \cos^2 \phi
 (1.73 + \tan \phi)^2~~,
\end{equation}
\begin{equation}
\Gamma(\ppp \to \gamma \chi_{c1}) = 176~{\rm keV} \cos^2 \phi
 (-0.87 + \tan \phi)^2~~,
\end{equation}
\begin{equation}
\Gamma(\ppp \to \gamma \chi_{c2}) = 167~{\rm keV} \cos^2 \phi
 (0.17 + \tan \phi)^2~~,
\end{equation}
and
\begin{equation}
\Gamma(\pp \to \gamma \chi_{c0}) = 67~{\rm keV} \cos^2 \phi
 (1 - 1.73 \tan \phi)^2~~,
\end{equation}
\begin{equation}
\Gamma(\pp \to \gamma \chi_{c1}) = 56~{\rm keV} \cos^2 \phi
 (1 + 0.87 \tan \phi)^2~~,
\end{equation}
\begin{equation}
\Gamma(\pp \to \gamma \chi_{c2}) = 39~{\rm keV} \cos^2 \phi
 (1 - 0.17 \tan \phi)^2~~.
\end{equation}
These quantities are plotted as functions of $\phi$ in Fig.\ \ref{fig:psimix}.

The observed ratio $R_{\ppp/\psi'}$ agrees with predictions only for
$\phi = (12 \pm 2)^\circ$ or $(-27 \pm 2)^\circ$, as shown by the vertical
bands in Fig.\ \ref{fig:psimix}.  Only the solution with $\phi = (12 \pm
2)^\circ$ is remotely consistent with the observed partial widths
\cite{PDG} $\Gamma(\psi' \to \gamma \chi_{cJ}) = 20$--30 keV.  This range of
$\phi$ favors the decay $\ppp \to \gamma \chi_{c0}$ over $\ppp \to \gamma
\chi_{c1,2}$ by a substantial amount.  The choice $\phi = (12 \pm 2)^\circ$
also is favored by the comparison of $\pp$ and $\ppp$ decays to $J/\psi
\pi^+ \pi^-$.  With the choice $\phi = (-27 \pm 2)^\circ$,
a larger rate would be predicted for $\ppp \to J/\psi \pi^+ \pi^-$ than for
$\pp \to J/\psi \pi^+ \pi^-$, in conflict with experiment \cite{Kuang:ub}.

Coupling to open $D \bar D$ channels and mixing schemes more general than Eq.\
(\ref{eqn:mix}) can affect radiative decay widths \cite{Eichten:2004uh}.  Table
\ref{tab:pw} compares partial widths predicted in one such scheme with those
based on Eq.\ (\ref{eqn:mix}).  In Ref.\ \cite{Eichten:2004uh} the
$\ppp$ is composed of only 52\% $ c \bar c$; the remainder of its wave
function contains additional light quark-antiquark pairs, {\it e.g.}, in the
form of the open $D \bar D$ channel.

\begin{table}
\caption{Partial widths in keV predicted in Ref.\ \cite{Eichten:2004uh}
without (a) or with (b) couplings to open channels and in Ref.\
\cite{Rosner:2001nm}.  $M(\ppp) = 3772$ MeV/$c^2$ is taken in accord
with the fit of Fig.\ 1; the nominal mass quoted in Ref.\ \cite{PDG} is
$3770.0 \pm 2.4$ MeV/$c^2$. \label{tab:pw}}
\begin{center}
\begin{tabular}{c c c c c} \hline \hline
$\psi''$& $E_\gamma$ & \multicolumn{2}{c}{Ref.\ \cite{Eichten:2004uh}} & Ref.\
\cite{Rosner:2001nm} \\
decay   &  (MeV)  & (a) & (b) & ($\phi = 12 \pm 2^\circ$) \\ \hline
$\gamma \chi_{c2}$ & 210 & 3.2 & 3.9 & $24 \pm 4$ \\
$\gamma \chi_{c1}$ & 252 & 183 & 59  & $73 \pm 9$ \\
$\gamma \chi_{c0}$ & 340 & 254 & 225 & $523 \pm 12$ \\ \hline \hline
\end{tabular}
\end{center}
\end{table}

The Mark III collaboration \cite{Zhu:1988} reported some marginal signals for
$\ppp$ radiative decays.  The prospects for observing $\ppp \to \gamma
\chi_{cJ}$ have been improved with the accumulation of $\int {\cal L} dt \simeq
56$ pb$^{-1}$ in the CLEO-c detector \cite{He:2005bs}.  With $\sigma(\ppp) \ge
6$ nb one should see several events in the cascade $\ppp \to \gamma \chi_{c1}
\to \gamma \gamma J/\psi \to \gamma \gamma \ell^+ \ell^-$. The inclusive signal
in $\ppp \to \gamma \chi_{c0}$ will not be statistics-limited.  To sum up, it
is unlikely that the total of the radiative widths $\Gamma(\psi'' \to \gamma
\chi_{cJ})$ exceeds about 600 keV, corresponding to a branching ratio slightly
above 2\%.

An early Mark III result \cite{Zhu:1988} found $\sigma(\ppp)
{\cal B}(\ppp \to J/\psi \pi^+ \pi^-) = (1.2 \pm 0.5 \pm 0.2) \times 10^{-2}$
nb, implying ${\cal B}(\ppp \to J/\psi \pi^+ \pi^-) = (0.15 \pm 0.07)\%$.
The BES Collaboration finds ${\cal B}(\ppp \to J/\psi \pi^+ \pi^-) = (0.34
\pm 0.14 \pm 0.08)\%$ \cite{Bai:2003hv}.  The average (not including
information from a CLEO upper limit \cite{Skwarnicki:2003wn} $< 0.26~(90\%~{\rm
c.l.})$) is ${\cal B}(\ppp \to J/\psi \pi^+ \pi^-) = (0.18 \pm 0.06)\%$,
corresponding to a partial width of $43 \pm 14$ keV.  Adding another 50\% for
$\ppp \to J/\psi \pi^0 \pi^0$, one finds $\Gamma(\ppp \to J/\psi \pi \pi)
= (64 \pm 21)$ keV, or at most about 100 keV.

At most 600 keV of the $\ppp$ total width of $23.6 \pm 2.7$ MeV is due to
radiative decays, and as much as another 100 keV is due to $J/\psi \pi
\pi$ decays.  Along with the predominant $D \bar D$ decays, these
contributions fall short of accounting for the total $\ppp$ width.

The total cross section for $e^+ e^- \to \ppp$ is not the only contribution to
hadron production at the $\ppp$ energy.  Continuum production from $e^+
e^- \to q \bar q$ $(q=u,d,s)$ should account for $\sigma(e^+ e^- \to q
\bar q) = 2 \sigma(e^+ e^- \to \mu^+ \mu^-) [ 1 + {\rm QCD~ correction}]$,
where $\sigma(e^+ e^- \to \mu^+ \mu^-) = 4 \pi \alpha^2/3s = 6.1~{\rm nb}$
for $s \equiv E_{\rm c.m.}^2 = (3770~{\rm MeV})^2$.  [Moreover $\tau^+ \tau^-$
pair production would account for $\sigma(e^+ e^- \to \tau^+ \tau^-) =( 1 - 4
m_\tau^2/s)^{1/2} ( 1 + 2 m_\tau^2/s) \sigma(e^+ e^- \to \mu^+ \mu^-) \simeq
2.9$ nb if initial-state-radiation effects were neglected.] The contribution of
the isovector photon $(G=+)$ dominates: $\sigma(2 \pi+ 4 \pi+ \ldots) = 9
\sigma (3 \pi+ 5 \pi+ \ldots)$.  Thus several even-G signatures of continuum
production can be examined at the $\ppp$ peak.  A better way to study continuum
contributions is to change the c.m.\ energy to one where resonance production
cannot contribute.  The CLEO Collaboration has done this, studying hadron
production at a c.m.\ energy of 3670 MeV with a sample of 21 pb$^{-1}$
\cite{Asner:2004yu}, and results are currently being analyzed.

Taking $\sigma(D \bar D) \le 6.5$ nb and comparing it
with the overall average of $\sigma(\ppp) = 7.9$ nb in Table \ref{tab:sig}, one
must account for a deficit of 1.4 nb, or 18\% of the total.  The possibilities
for detecting {\it individual} charmless decay modes of the $\ppp$ were raised,
for example, in Refs.\ \cite{Rosner:2001nm,Wang:2004kf,Achasov:vh}.  Here I
stress that more inclusive measurements at the $\ppp$ may be of use.

Consider a model in which the re-annihilation of charmed quarks in $D^0 \bar
D^0$ and $D^+ D^-$ into states containing $u,~d,~s$ accounts for the difference
between $\sigma(D \bar D)$ and $\sigma(\ppp)$.  Such re-annihilation was
proposed \cite{Lipkin:1986av} both as a source of non-$D \bar D$ decays of the
$\ppp$ and as a possible source of non-$B \bar B$ decays of the $\Upsilon(4S)$.
The latter do not occur at any level above a few percent \cite{nonBB}.

The BES Collaboration's continuum value $R = 2.26 \pm 0.14$ (from a preliminary
version of \cite{Ablikim:2004ck})
averaged over 2 GeV $ \le E_{\rm c.m.} \le$ 3 GeV is consistent with the
expected value of 2 times a QCD correction and with the background obtained in
the fit of Fig.\ \ref{fig:sigppp} to the $\ppp$ cross section. I take $R=2.26$.
Of this, one expects $R(s \bar s) = (1/6)(2.26) = 0.377$.
The non-strange contributions may be decomposed into a 9:1 ratio of $I=1$ and
$I=0$ contributions denoted by $R_1$ and $R_0$, since $(Q_u - Q_d)^2 =
9(Q_u + Q_d)^2$.  Thus $R_1 = (5/6)(2.26)(9/10) = 1.695$ and $R_0 =
(5/6)(2.26)(1/10) = 0.188$.  The $I=1$ continuum corresponds to an isovector
photon and even-G-parity states, while the $s \bar s$ and $I=0$ nonstrange
continuua correspond to an isoscalar photon and odd-G-parity states.  The $s
\bar s$ continuum is expected to yield
at least one $K \bar K$ pair in its hadronic products.

I take the amplitude for $\ppp \to D \bar D \to$ (non-charmed final states)
to proceed via a $D \bar D$ loop diagram
characterized by an amplitude proportional to ${p^*}^3$, where $p^*$ is the
magnitude of the c.m.\ 3-momentum of either $D$.  For $\ppp \to D^+ D^-$,
$p^*_{+-} = 250.0$ MeV/$c$, while for $\ppp \to D^0 \od$, $p^*_{00} = 283.6$
MeV/$c$.  The re-annihilation amplitude $A^R_d$ into $d \bar d$ pairs and
the amplitude $A^R_u$ into $u \bar u$ pairs are then expected to be in the
ratio $A^R_d/A^R_u = (p_{+-}^*/p^*_{00})^3 = 0.685$, and the corresponding
ratio for isovector and nonstrange isoscalar contributions $A^R_1$ and $A^R_0$
is
\begin{equation}
\frac{A^R_1}{A^R_0} = \frac{A^R_u-A^R_d}{A^R_u+A^R_d}
= \frac{1 - 0.685}{1 + 0.685} = 0.187~~.
\end{equation}

I assume that the re-annihilation amplitudes into $I=0$ and $I=1$ final states
have the same strong phase $\delta$ relative to the continuum, modulated by a
Breit-Wigner amplitude $f_B$ defined to be unity at the resonance peak, since
the $I=1$ contribution arises largely from the mass difference between charged
and neutral $D$ mesons but in other respects arises from the same rescattering
mechanism (charm-anticharm annihilation to light non-strange quarks) as the
$I=0$ contribution.  This assumption may be relaxed if the interference
patterns to be discussed below are found to differ for $I=0$ and $I=1$ channels.

In the vicinity of the $\ppp$ mass $M_0$ one may
then write the amplitudes $A_1$ and $A_0$ for the isovector and nonstrange
isoscalar contributions to $R$ as functions of c.m.\ energy $E$:
\begin{equation}
A_1 = 0.187 b_0 e^{i \delta} f_B(E) + \sqrt{R_1}~~,~~~
A_0 = b_0 e^{i \delta} f_B(E) + \sqrt{R_0}~~,
\end{equation}
where the amplitudes have been defined such that their squares yield
their contributions to $R$, and
\begin{equation}
f_B(E) = [d_B(E)]^{-1}~~,~~~d_B(E) \equiv 1 + \frac{2 i (M_0 - E)}{\Gamma}~~.
\end{equation}
The values $M_0 = 3772$ MeV/$c^2$ and $\Gamma = 23.2$ MeV are taken from the
fit of Fig.\ \ref{fig:sigppp}.  This fit implies a peak value $R(M_0) = 3.53$
which will be taken as a constraint when choosing the arbitrary constant $b_0$.

The continuum away from the peak accounts for $R=2.26$, so one must provide
a total resonant contribution of $\Delta R_{\rm pk} = 3.53 - 2.26 = 1.27$.
Consider $D \bar D$ pairs to provide 82\% of this value,
or $\Delta R_{\rm pk}^{D \bar D} = 1.04$.  This contribution will be modulated
by $|f_B(E)|^2$.  There will be a constant $s \bar s$ continuum contribution
of $\Delta R^{s \bar s} = 0.38$, and contributions from the isovector and
non-strange isoscalar amplitudes $A_I$ above, leading to a total of
\begin{equation}
R(E) = |A_1|^2 + |A_0|^2 + \Delta R_{\rm pk}^{D \bar D}|f_B(E)|^2 +
\Delta R^{s \bar s}~~.
\end{equation}

\begin{figure}
\includegraphics[height=6.2in]{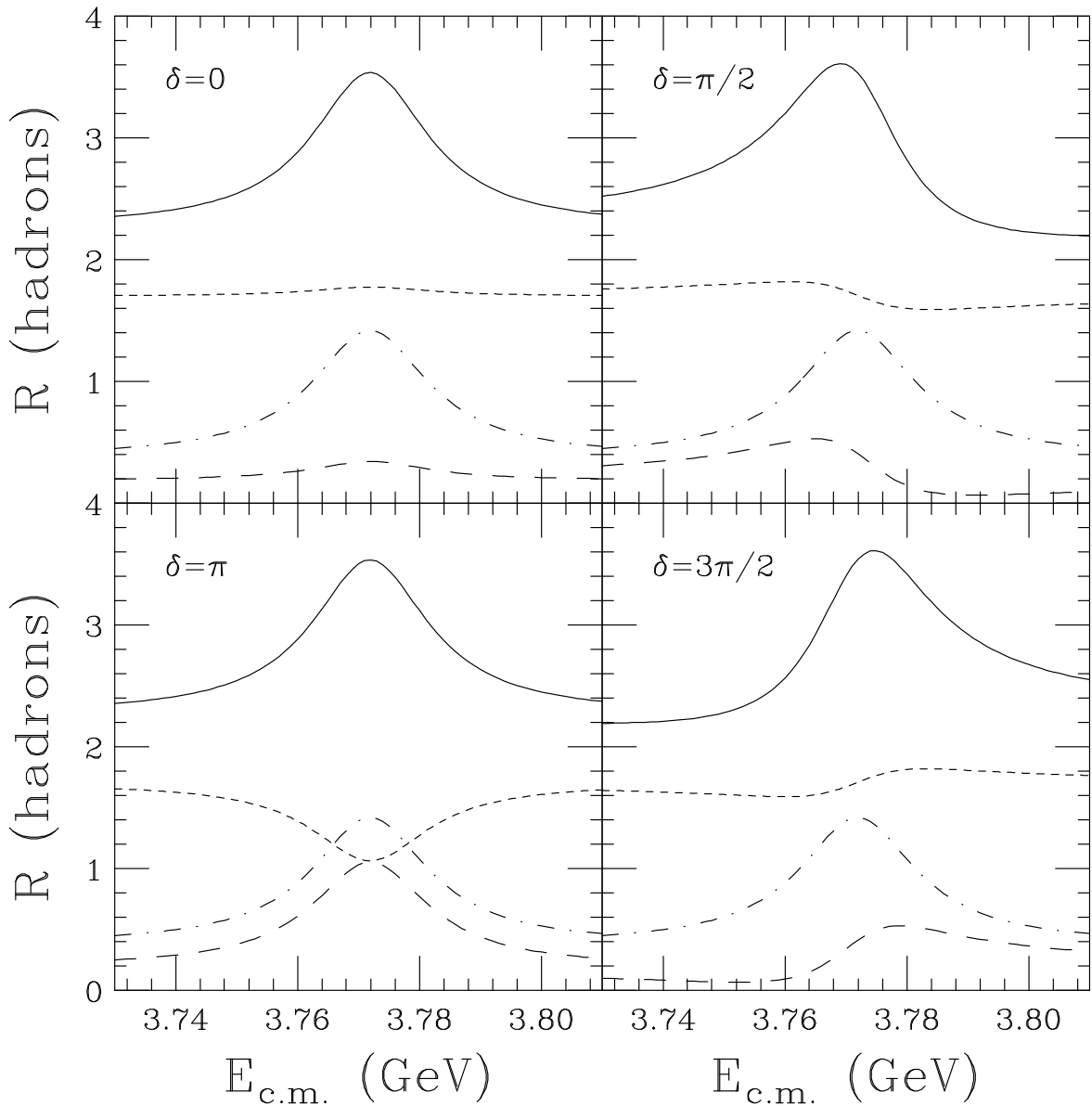}
\caption{Contributions to $R$ in the vicinity of the $\ppp$ resonance energy.
Solid curves: total, constrained to have a value of 3.53 at $M(\ppp) = 3.772$
GeV/$c^2$.
Short-dashed curves: $I=1$ continuum interfering with $I=1$ contribution from
$D \bar D$ reannihilation.  Long-dashed curves: $I=0$ non-strange continuum
interfering with $I=0$ nonstrange contribution from $D \bar D$ reannihilation.
Dot-dashed curves:  $D \bar D$ resonance contribution, taken to contribute
82\% of resonance peak cross section, plus $s \bar s$ continuum.
\label{fig:rall}}
\end{figure}

For $\delta = 0$, a modest value $b_0 = 0.15$ provides
the additional contribution needed to account for the missing 18\% of the
$\ppp$ peak cross section.  The corresponding values for $\delta = \pi/2,
\pi,3\pi/2$ are 0.47, 1.46, and 0.47, respectively.
Fig.\ \ref{fig:rall} displays the result of this calculation, in which
re-annihilation accounts for 18\% of the peak $R$ value at $M(\ppp) =
3772$ MeV/$c^2$.  A relative phase $\delta$ between the reannihilation
amplitude and the continuum was defined in such a way that $\delta = 0$
corresponds to constructive interference at the resonance peak.
Several features of this model are worth noting.

\begin{itemize}

\item The re-annihilation of $D^+ D^-$ and $D^0 \bar D^0$ pairs into light
quarks will favor leading $ d \bar d$ and $u \bar u$ pairs, with amplitudes
in the ratio $d \bar d : u \bar u \simeq 2:3$ in line with the cross section
ratio $\sigma(D^+ D^-):\sigma(D^0 \bar D^0)$.  The
fragmentation of these quarks will populate hadronic final states in
different proportions than the usual continuum process in which quark pairs are
produced by the virtual photon with amplitudes proportional to their charges.

\item The re-annihilation favors isoscalar ($I=0$) odd-G-parity
final states, so one should see more effects of interference between
re-annihilation and continuum in odd G ($3 \pi,5 \pi, \eta 3 \pi, \eta' 3 \pi,
\ldots$) states than in even-G ones ($2 \pi, 4 \pi, \eta 2 \pi, \ldots$).
This interference is particularly pronounced because the larger odd-G
reannihilation amplitude is interfering with a smaller odd-G continuum
amplitude.

\item The effects of re-annihilation on the continuum contributions will be
hard to see if $\delta = 0$, especially in the dominant $I=1$ (even-G-parity)
channel.  They are proportionately greater in the $I=0$ (odd-G-parity)
non-strange channel (consisting, for example, of odd numbers of pions).

\item The re-annihilation may be similar to that which accounts for enhanced
penguin contributions in $B$ decays, particularly in the $b \to s$ subprocess
through the chain $b \to c \bar c s \to q \bar q s$, where $q=(u,d,s)$
(see also \cite{Rosner:2001nm}).  If this is so, one should look for an
enhancement of $\eta'$ production as occurs in inclusive and
exclusive $B$ decays.

\item As evident for non-zero $\delta$, measurement
of the cross section in semi-inclusive channels with definite G-parity
and especially odd G (such as final states with an odd number of pions) may
show interesting interference patterns over an energy range $M(\ppp) \pm
\Gamma(\ppp)/2 \simeq 3772 \pm 12$ MeV/$c^2$.

\end{itemize}

A Breit-Wigner amplitude is normally taken to be purely imaginary at its peak.
I incorporate this phase into the definition of $\delta$.  The choice $\delta =
3 \pi/2$ would correspond to no additional phase associated with
the re-annihilation process, for example in $e^+ e^- \to \mu^+ \mu^-$ in the
vicinity of the resonance, where interference between continuum and resonance
is destructive below the resonance and constructive above it.  (For an example
of this behavior at the $\psi'$, see Ref.\ \cite{Wang:2004qg}.)
It was speculated in Refs.\ \cite{Rosner:2001nm}
and \cite{Suzuki:fs} (see also Refs.\ \cite{Rosner:1999zm})
that such an additional phase could be present and,
if related to a similar phase in $B$ decays, might account for a strong
phase in penguin $b \to s$ amplitudes.  A recent fit to $B \to P P$ decays,
where $P$ denotes a charmless pseudoscalar meson \cite{Chiang:2004nm}, finds
such a phase to be in the range of roughly $-20^\circ$ to $-50^\circ$.  This
would correspond to taking $\delta$ in the range of $40^\circ$ to $70^\circ$.
The presence of such a phase is supported by the recent strengthening of
the evidence for a significant CP asymmetry in the decay $B^0 \to K^+ \pi^-$
\cite{Aubert:2004qm}.

I now discuss briefly some exclusive charmless decay modes of the $\ppp$.  It
was suggested in Ref.\ \cite{Rosner:2001nm} that some $\pp$ decay modes
might be suppressed via S--D mixing.  In that case, they should show up
in $\ppp$ decays.  Foremost among these was the $\pp \to \rho \pi$ decay.
It was then pointed out \cite{Wang:2003zx} that because of possible
interference with continuum, decays such as $\ppp \to \rho \pi$ might
manifest themselves in various ways depending on relative strong phases, even
as a {\it dip} in $\sigma(e^+ e^- \to \rho \pi)$ at $M(\ppp)$.

All of the suppressed $\ppp$ modes discussed in Refs.\ \cite{Rosner:2001nm} and
\cite{Bai:2002yn} are prime candidates for detection in $\ppp$ decays.  The
interference proposed in Ref.\ \cite{Wang:2003zx} can actually lead to a {\it
suppression} of some modes relative to the rate expected from continuum. It was
anticipated in Ref.\ \cite{Rosner:2001nm} that if one were to account for any
``missing'' $\pp$ decay modes by mixing with the $\ppp$, such an effect need
not contribute more than a percent or two to the total $\ppp$ width.  However,
Ref.\ \cite{Wang:2004kf} recently obtained a charmless $\ppp$ branching ratio
of up to 13\% obtained by generalizing the above arguments to all charmless
final states of $\psi'$ and $\ppp$ within the S--D mixing framework.

To sum up:

(1) Some non--$D \bar D$ decay modes of the $\psi''$ do exist, such as $\ell^+
\ell^-$ pairs, $\gamma \chi_{cJ}$, and $J/\psi \pi \pi$.  They tell us about
mixing between S-waves, D-waves, and open $D \bar D$ channels.

(2) Most non--$D \bar D$ final states at the $\psi''$ are from continuum
production.  Their yields will not vary much with beam energy unless their
continuum production amplitudes are interfering with a genuine Breit-Wigner
contribution from the $\psi''$.  This interference is most likely to show up in
odd-G-parity final states, for which appreciable distortions of the
Breit-Wigner line shape can occur.

(3) I predict a substantial enhancement of $\eta'$ production in charmless
$\ppp$ final states if the re-annihilation of $D \bar D$ into light quarks is
related to the generation of a $b \to s$ penguin amplitude in $B$ decays.

(4) The suggestion that the ``missing'' $\psi'$ decays, like $\rho \pi$, should
show up instead at the $\psi''$, is being realized, if at all, in a more subtle
manner, and does not illuminate the question of whether a substantial fraction
(at least several percent) of the $\psi''$ cross section is non--$D \bar D$.

(5) The measurement of the continuum cross section at 3670 MeV is expected to
yield $R = 2 (1 + \alpha_S/\pi + \ldots)$.  Its value, when extrapolated to
3770 MeV, is relevant to whether there is a cross section deficit at the
$\psi''$.

(6) Proposed experimental tests of these points include (a) a scan of the
$\ppp$ peak to measure $\sigma(\ppp)$ more accurately, with an eye to
the possibility of different behavior in different channels and distortion of
the peak shape due to resonant-continuum interference; (b) reduction
of the error on $\sigma(D \bar D)$; and (c) use of continuum data to perform
an analysis of the total hadron production cross section at the $\ppp$ energy.

I thank Karl Berkelman, David Cassel, Daniel Cronin-Hennessy, Richard Galik,
Brian Heltsley, Hanna Mahlke-Kr\"uger, Hajime Muramatsu, Ian Shipsey, Tomasz
Skwarnicki, and Misha Voloshin for discussions, and Maury Tigner for extending
the hospitality of the Laboratory for Elementary-Particle Physics at Cornell
during this research.  This work was supported in part by the United States
Department of Energy under Grant No.\ DE FG02 90ER40560, by the National
Science Foundation under Grant No.\ 0202078, and by the John Simon Guggenheim
Memorial Foundation.

\def \ajp#1#2#3{Am.\ J. Phys.\ {\bf#1}, #2 (#3)}
\def \apny#1#2#3{Ann.\ Phys.\ (N.Y.) {\bf#1}, #2 (#3)}
\def \app#1#2#3{Acta Phys.\ Polonica {\bf#1}, #2 (#3)}
\def \arnps#1#2#3{Ann.\ Rev.\ Nucl.\ Part.\ Sci.\ {\bf#1}, #2 (#3)}
\def \b97{{\it Beauty '97}, Proceedings of the Fifth International
Workshop on $B$-Physics at Hadron Machines, Los Angeles, October 13--17,
1997, edited by P. Schlein}
\def \art{and references therein}
\def \cmts#1#2#3{Comments on Nucl.\ Part.\ Phys.\ {\bf#1}, #2 (#3)}
\def \cn{Collaboration}
\def \cp89{{\it CP Violation,} edited by C. Jarlskog (World Scientific,
Singapore, 1989)}
\def \ctp#1#2#3{Commun.\ Theor.\ Phys.\ {\bf#1}, #2 (#3)}
\def \efi{Enrico Fermi Institute Report No.\ }
\def \epjc#1#2#3{Eur.\ Phys.\ J. C #1 (#3) #2}
\def \f79{{\it Proceedings of the 1979 International Symposium on Lepton and
Photon Interactions at High Energies,} Fermilab, August 23-29, 1979, ed. by
T. B. W. Kirk and H. D. I. Abarbanel (Fermi National Accelerator Laboratory,
Batavia, IL, 1979}
\def \hb87{{\it Proceeding of the 1987 International Symposium on Lepton and
Photon Interactions at High Energies,} Hamburg, 1987, ed. by W. Bartel
and R. R\"uckl (Nucl.\ Phys.\ B, Proc.\ Suppl., vol.\ 3) (North-Holland,
Amsterdam, 1988)}
\def \ib{{\it ibid.}~}
\def \ibj#1#2#3{~#1 (#3) #2}
\def \ichep72{{\it Proceedings of the XVI International Conference on High
Energy Physics}, Chicago and Batavia, Illinois, Sept. 6 -- 13, 1972,
edited by J. D. Jackson, A. Roberts, and R. Donaldson (Fermilab, Batavia,
IL, 1972)}
\def \ijmpa#1#2#3{Int.\ J.\ Mod.\ Phys.\ A {\bf#1}, #2 (#3)}
\def \ite{{\it et al.}}
\def \jhep#1#2#3{JHEP {\bf#1}, #2 (#3)}
\def \jpb#1#2#3{J.\ Phys.\ B {\bf#1}, #2 (#3)}
\def \lg{{\it Proceedings of the XIXth International Symposium on
Lepton and Photon Interactions,} Stanford, California, August 9--14 1999,
edited by J. Jaros and M. Peskin (World Scientific, Singapore, 2000)}
\def \lkl87{{\it Selected Topics in Electroweak Interactions} (Proceedings of
the Second Lake Louise Institute on New Frontiers in Particle Physics, 15 --
21 February, 1987), edited by J. M. Cameron \ite~(World Scientific, Singapore,
1987)}
\def \kdvs#1#2#3{{Kong.\ Danske Vid.\ Selsk., Matt-fys.\ Medd.} {\bf #1},
No.\ #2 (#3)}
\def \ky85{{\it Proceedings of the International Symposium on Lepton and
Photon Interactions at High Energy,} Kyoto, Aug.~19-24, 1985, edited by M.
Konuma and K. Takahashi (Kyoto Univ., Kyoto, 1985)}
\def \mpla#1#2#3{Mod.\ Phys.\ Lett.\ A {\bf#1}, #2 (#3)}
\def \nat#1#2#3{Nature {\bf#1}, #2 (#3)}
\def \nc#1#2#3{Nuovo Cim.\ {\bf#1}, #2 (#3)}
\def \nima#1#2#3{Nucl.\ Instr.\ Meth. A #1 (#3) #2}
\def \np#1#2#3{Nucl.\ Phys.\ #1 (#3) #2}
\def \npbps#1#2#3{Nucl.\ Phys.\ B Proc.\ Suppl.\ {\bf#1}, #2 (#3)}
\def \os{XXX International Conference on High Energy Physics, Osaka, Japan,
July 27 -- August 2, 2000}
\def \PDG{Particle Data Group, S. Eidelman et al., \plb {592}{1}{2004}}
\def \pisma#1#2#3#4{Pis'ma Zh.\ Eksp.\ Teor.\ Fiz.\ {\bf#1}, #2 (#3) [JETP
Lett.\ {\bf#1}, #4 (#3)]}
\def \pl#1#2#3{Phys.\ Lett.\ {\bf#1}, #2 (#3)}
\def \pla#1#2#3{Phys.\ Lett.\ A {\bf#1}, #2 (#3)}
\def \plb#1#2#3{Phys.\ Lett.\ B #1 (#3) #2}
\def \pr#1#2#3{Phys.\ Rev.\ {\bf#1}, #2 (#3)}
\def \prc#1#2#3{Phys.\ Rev.\ C {\bf#1}, #2 (#3)}
\def \prd#1#2#3{Phys.\ Rev.\ D #1 (#3) #2}
\def \prl#1#2#3{Phys.\ Rev.\ Lett.\ #1 (#3) #2}
\def \prp#1#2#3{Phys.\ Rep.\ {\bf#1}, #2 (#3)}
\def \ptp#1#2#3{Prog.\ Theor.\ Phys.\ {\bf#1}, #2 (#3)}
\def \rmp#1#2#3{Rev.\ Mod.\ Phys.\ {\bf#1}, #2 (#3)}
\def \rp#1{~~~~~\ldots\ldots{\rm rp~}{#1}~~~~~}
\def \rpp#1#2#3{Rep.\ Prog.\ Phys.\ {\bf#1}, #2 (#3)}
\def \sing{{\it Proceedings of the 25th International Conference on High Energy
Physics, Singapore, Aug. 2--8, 1990}, edited by. K. K. Phua and Y. Yamaguchi
(Southeast Asia Physics Association, 1991)}
\def \slc87{{\it Proceedings of the Salt Lake City Meeting} (Division of
Particles and Fields, American Physical Society, Salt Lake City, Utah, 1987),
ed. by C. DeTar and J. S. Ball (World Scientific, Singapore, 1987)}
\def \slac89{{\it Proceedings of the XIVth International Symposium on
Lepton and Photon Interactions,} Stanford, California, 1989, edited by M.
Riordan (World Scientific, Singapore, 1990)}
\def \smass82{{\it Proceedings of the 1982 DPF Summer Study on Elementary
Particle Physics and Future Facilities}, Snowmass, Colorado, edited by R.
Donaldson, R. Gustafson, and F. Paige (World Scientific, Singapore, 1982)}
\def \smass90{{\it Research Directions for the Decade} (Proceedings of the
1990 Summer Study on High Energy Physics, June 25--July 13, Snowmass, Colorado),
edited by E. L. Berger (World Scientific, Singapore, 1992)}
\def \tasi{{\it Testing the Standard Model} (Proceedings of the 1990
Theoretical Advanced Study Institute in Elementary Particle Physics, Boulder,
Colorado, 3--27 June, 1990), edited by M. Cveti\v{c} and P. Langacker
(World Scientific, Singapore, 1991)}
\def \yaf#1#2#3#4{Yad.\ Fiz.\ {\bf#1}, #2 (#3) [Sov.\ J.\ Nucl.\ Phys.\
{\bf #1}, #4 (#3)]}
\def \zhetf#1#2#3#4#5#6{Zh.\ Eksp.\ Teor.\ Fiz.\ {\bf #1}, #2 (#3) [Sov.\
Phys.\ - JETP {\bf #4}, #5 (#6)]}
\def \zpc#1#2#3{Zeit.\ Phys.\ C {\bf#1}, #2 (#3)}
\def \zpd#1#2#3{Zeit.\ Phys.\ D {\bf#1}, #2 (#3)}


\begin{thebibliography}{99}

\bibitem{CLEOc} CLEO Collaboration, ``CLEO-c and CESR-c: A New Frontier of
Weak and Strong Interactions,'' 2001 (unpublished).  See
{\tt http://www.lns.cornell.edu/public/CLEO/spoke/CLEOc/}
for a description of plans
for running of CLEO/CESR at the $\ppp$ and other energies.

\bibitem{Ablikim:2004ck}
  M.~Ablikim et al.\  [BES Collaboration],
  Phys.\ Lett.\ B 603 (2004) 130.

\bibitem{Eichten:2002qv}
E.~J.~Eichten, K.~Lane and C.~Quigg,
Phys.\ Rev.\ Lett.\ 89 (2002) 162002.

\bibitem{Eichten:2004uh}
E.~J.~Eichten, K.~Lane and C.~Quigg,
Phys.\ Rev.\ D 69 (2004) 094019.

\bibitem{Rosner:2001nm}
J.~L.~Rosner,
Phys.\ Rev.\ D 64 (2001) 094002.

\bibitem{Wang:2003hy}
P.~Wang, C.~Z.~Yuan and X.~H.~Mo,
Phys.\ Rev.\ D 69 (2004) 057502;
C.~Z.~Yuan, P.~Wang and X.~H.~Mo,
Phys.\ Lett.\ B 567 (2003) 73;
P.~Wang, C.~Z.~Yuan and X.~H.~Mo,
Phys.\ Lett.\ B 574 (2003) 41;
P.~Wang, X.~H.~Mo and C.~Z.~Yuan, Phys.\ Rev.\ D 70 (2004) 077505;
P.~Wang,
arXiv:hep-ph/0410028.

\bibitem{Wang:2004kf}
P.~Wang, C.~Z.~Yuan and X.~H.~Mo, Phys.\ Rev.\ D 70 (2004) 114014.

\bibitem{He:2005bs}
  Q.~He {\it et al.}  [CLEO Collaboration],
  arXiv:hep-ex/0504003.

\bibitem{Adler:1987as}
J.~Adler {\it et al.}  [MARK-III Collaboration],
Phys.\ Rev.\ Lett.\  60 (1988) 89.

\bibitem{Partridge:1984kb}
R.~A.~Partridge [Crystal Ball Collaboration],
``A Study Of The $\psi''(3770)$ Using The Crystal Ball Detector,''
Ph.~D. Thesis, California Institute of Technology, 1984, Caltech
Report No.\ CALT-68-1150 (unpublished).

\bibitem{Peruzzi:1977ms}
I.~Peruzzi {\it et al.} [Lead-Glass Wall Collaboration],
Phys.\ Rev.\ Lett.\ 39 (1977) 1301;
D.~L.~Scharre {\it et al.},
Phys.\ Rev.\ Lett.\ 40 (1978) 74.

\bibitem{Schindler:1980ws}
R.~H.~Schindler {\it et al.} [Mark II Collaboration],
Phys.\ Rev.\ D 24 (1981) 78.

\bibitem{Bai:2001ct}
J.~Z.~Bai {\it et al.}  [BES Collaboration],
Phys.\ Rev.\ Lett.\ 88 (2002) 101802.

\bibitem{Lipkin:1986av}
H.~J.~Lipkin,
Phys.\ Lett.\ B 179 (1986) 278.

\bibitem{Achasov:vh}
N.~N.~Achasov and A.~A.~Kozhevnikov,
Phys.\ Rev.\ D 49 (1994) 275.

\bibitem{Achasov:1999tp}
N.~N.~Achasov and A.~A.~Kozhevnikov,
Phys.\ Rev.\ D 61 (2000) 054005.

\bibitem{Zhu:1988} Yanong Zhu, Ph.\ D. Thesis, California Institute of
Technology, 1988, Caltech report CALT-68-1513 (unpublished).

\bibitem{Majid:1993} Walid Abdul Majid, Ph.\ D. Thesis, University of Illinois,
1993 (unpublished).

\bibitem{PDG} \PDG.

\bibitem{Voloshin:2004nu}
M.~B.~Voloshin,
arXiv:hep-ph/0402171.

\bibitem{BW} J. M. Blatt and V. F. Weisskopf, {\it Theoretical Nuclear
Physics}, Wiley, New York, 1952, pp.\ 358--365;
F.~Von Hippel and C.~Quigg,
Phys.\ Rev.\ D 5 (1972) 624.

\bibitem{TSPC} T. Skwarnicki (private communication), performing the same fit
with a different background parametrization, obtains $8.5 \pm 0.7$ nb.

\bibitem{Kuang:2002hz}
Y.~P.~Kuang,
Phys.\ Rev.\ D 65 (2002) 094024.

\bibitem{Kuang:ub}
Y.~P.~Kuang and T.~M.~Yan,
Phys.\ Rev.\ D 41 (1990) 155.

\bibitem{Bai:2003hv}
J.~Z.~Bai {\it et al.}  [BES Collaboration], Phys.\ Lett.\ B 605 (2005) 63.

\bibitem{Skwarnicki:2003wn}
T.~Skwarnicki,
Int.\ J.\ Mod.\ Phys.\ A 19 (2004) 1030.

\bibitem{Asner:2004yu}
D.~Asner,
``The CLEO-c Research Program,'' Cornell University Report No.\ CLNS 04/1875,
arXiv:hep-ex/0405009, presented at XXXIXth Rencontres de Moriond:  Electroweak
Interactions and Unified Theories, March 21--28, 2004, La Thuile, Italy.

\bibitem{nonBB} Non-$B \bar B$ decays of the $\Upsilon(4S)$ are discussed 
in Ref.\ \cite{Achasov:vh} and by J. L. Rosner, in {\it Research
Directions for the Decade} (Proceedings of the 1990 Summer Study on High Energy
Physics, June 25 -- July 13, 1990, Snowmass, Colorado), edited by E. L. Berger
(World Scientific, Singapore, 1992), p.\ 268.  Experimental upper limits
have not been published.

\bibitem{Wang:2004qg}
P.~Wang, in Hadron 03: 10th International Conference on Hadron Spectroscopy,
Aschaffenburg, Germany, 2003, AIP Conference Proceedings 717 (2004) 571.

\bibitem{Suzuki:fs}
M.~Suzuki,
Phys.\ Rev.\ D 63 (2001) 054021.

\bibitem{Rosner:1999zm}
J.~L.~Rosner,
Phys.\ Rev.\ D 60 (1999) 074029;
I. Dunietz, J. Incandela, F. D. Snider, and H. Yamamoto,
\epjc{1}{211}{1998}; I. Dunietz, in \b97, \nima{408}{14}{1998}, \art;
M. Ciuchini, E. Franco, G. Martinelli, and L. Silvestrini,
\np{B501}{271}{1997}; M. Ciuchini, R. Contino, E. Franco, G. Martinelli, and L.
Silvestrini, \np{B512}{3}{1998};
Y.-Y. Keum, H.-N. Li, and A. I. Sanda, \plb{504}{6}{2001};
\prd{63}{054008}{2001}; Y.-Y. Keum and H.-N. Li, \prd{63}{074006}{2001}.

\bibitem{Chiang:2004nm}
C.~W.~Chiang, M.~Gronau, J.~L.~Rosner and D.~A.~Suprun,
Phys.\ Rev.\ D 70 (2004) 034020.

\bibitem{Aubert:2004qm}
B.~Aubert {\it et al.}  [BaBar Collaboration],
Phys.\ Rev.\ Lett.\ 93 (2004) 131801:
$A(K^+ \pi^-) = -0.133 \pm 0.030 \pm 0.009$;
Y.~Chao [Belle Collaboration], presented at ICHEP04 (International Conference
on High Energy Physics, Beijing, 2004):
$A(K^+ \pi^-) = -0.101 \pm 0.025 \pm 0.005$.

\bibitem{Wang:2003zx}
P.~Wang, C.~Z.~Yuan and X.~H.~Mo,
Phys.\ Lett.\ B 574 (2003) 41.


\bibitem{Bai:2002yn}
J.~Z.~Bai {\it et al.}  [BES Collaboration],
Phys.\ Rev.\ D 67 (2003) 052002;
J.~Z.~Bai {\it et al.}  [BES Collaboration],
Phys.\ Rev.\ Lett.\ 92 (2004) 052001;
J.~Z.~Bai {\it et al.}  [BES Collaboration],
Phys.\ Rev.\ D 69 (2004) 072001;
N.~E.~Adam {\it et al.}  [CLEO Collaboration],
Phys.\ Rev.\ Lett.\ 94 (2005) 012005;
M.~Ablikim {\it et al.}  [BES Collaboration],
arXiv:hep-ex/0407037;
M.~Ablikim {\it et al.}  [BES Collaboration],
arXiv:hep-ex/0408047;
M.~Ablikim {\it et al.}  [BES Collaboration],
Phys.\ Rev.\ D 70 (2004) 112003;
M.~Ablikim {\it et al.}  [BES Collaboration],
Phys.\ Rev.\ D 70 (2004) 112007; ibid.\ 71 (2005) 019901 [Erratum].

\end{thebibliography}
\end{document}